\documentclass[preprint,12pt,3p]{elsarticle}

\usepackage{amssymb}
\usepackage{todonotes}
\usepackage{marvosym}
\usepackage{setspace}
\usepackage{threeparttable}

\begin{document}

\begin{frontmatter}

\title{ 
 Direct Numerical Simulation of Turbulent Channel Flow with Riblets at Low Reynolds Number}


\author[label1]{V. Fink}
\address[label1]{Institute of Fluid Mechanics, Karlsruhe Institute of Technology, Germany}

\author[label1]{B. Frohnapfel\corref{cor1}}
\cortext[cor1]{{bettina.frohnapfel@kit.edu}}

\end{frontmatter}

In this report we briefly summarize the set-up and results for direct numerical simulations (DNS) of turbulent channel flows with riblets. The DNS are carried out with OpenFOAM$\tiny{^{\textregistered}}$ which allows the use of an unstructured mesh. The present results were presented at the European Drag Reduction and Flow Control Meeting in 2015 \cite{finkexperimental}. 

\section{Procedure}
\label{sec1}

Due to the well known fact, that the sharp riblet tips are important to achieve drag reduction, we aim at a very high grid resolution in this area. The chosen riblet geometries are, first, the one investigated by Choi et al. \cite{choi1993direct} and, second, another trapezoidal geometry for which detailed experimental studies were carried out in our lab \cite{guttler2015high}. This latter one is similar to the riblet geometries investigated by Bechert and co-workers in the Berlin oil tunnel \cite{bechert1997experiments}.

\begin{figure}[h!]
\centering
\includegraphics[width=0.8\textwidth]{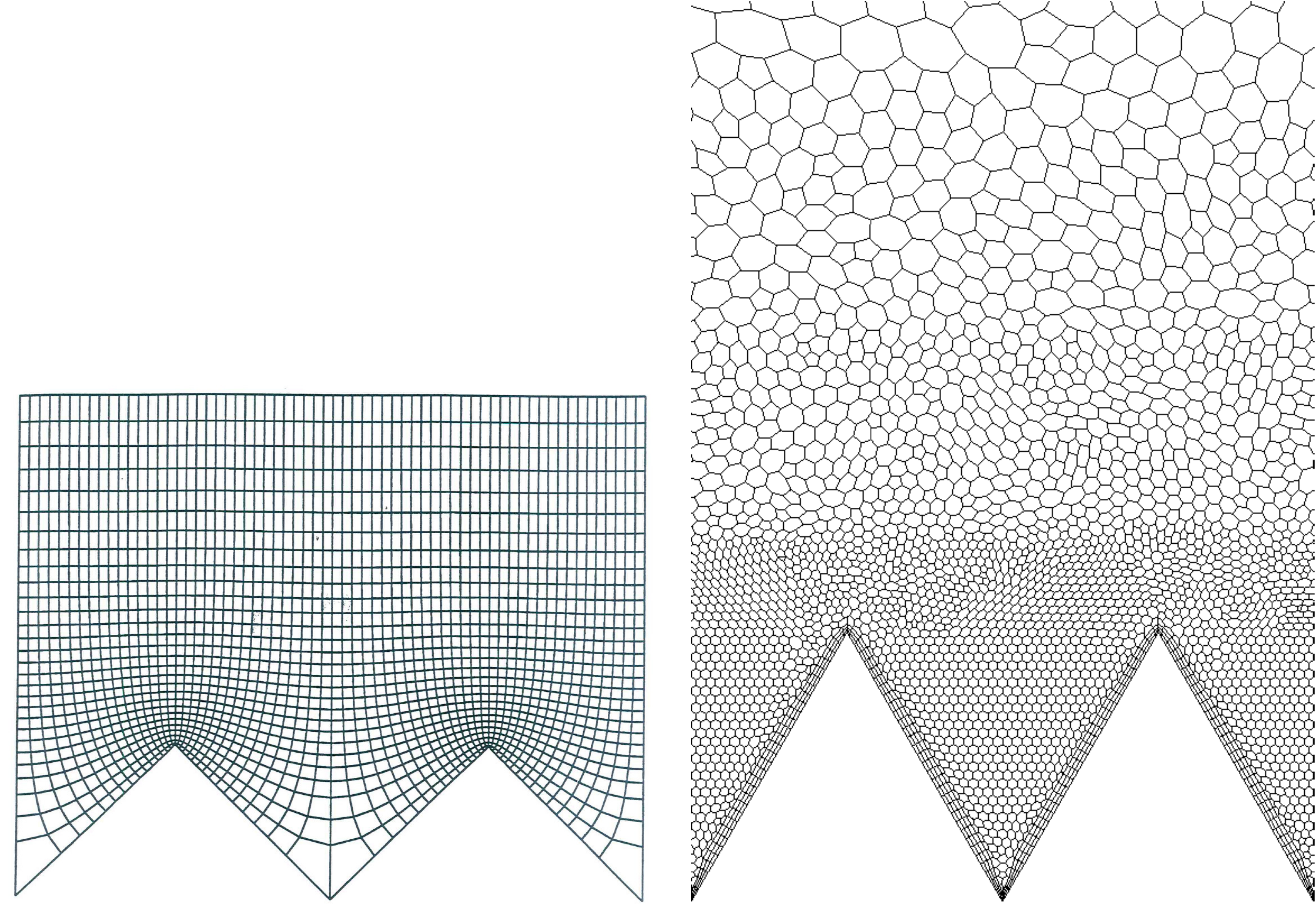}
\caption{(Left) Grid of Choi's case \cite{choi1993direct}. (Right) Grid of the present simulation.}
\label{Choi}
\end{figure}

\begin{figure}[h!]
\centering
\includegraphics[width=0.9\textwidth]{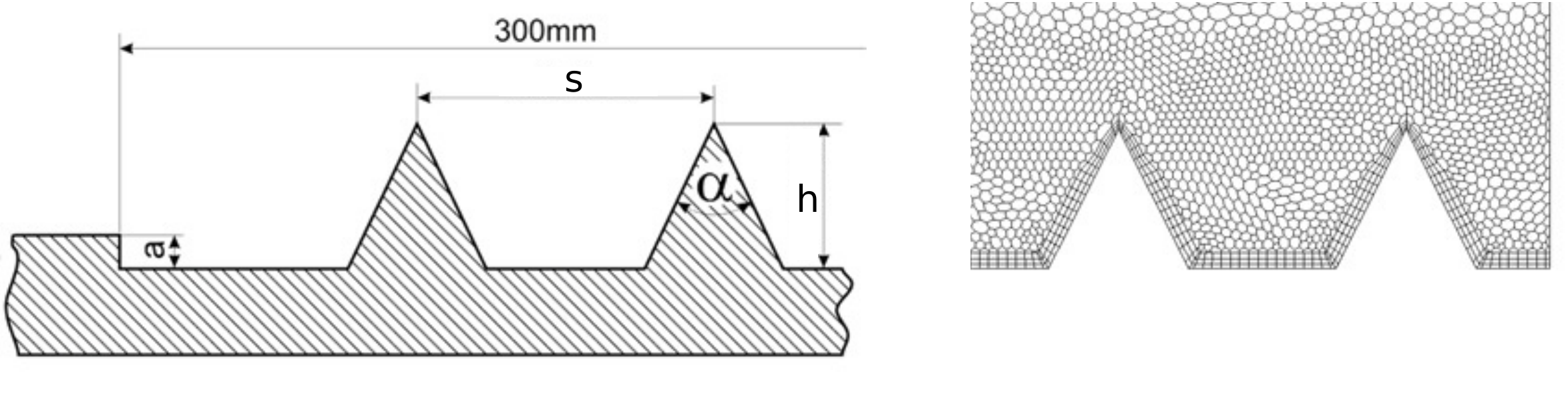}
\caption{ A section of the experimentally investigated trapezoidal riblet structure \cite{guttler2015high} on the left ($s=614\mu m$, $h=294\mu m$, $\alpha=53.5^\circ $, $a=70\mu m$) and the corresponding unstructured polyhedral grid in the vicinity of the riblets on the right side.}
\label{ribs}
\end{figure}

The DNS are carried out for a turbulent channel flow at a bulk Reynolds number of $Re_b=U_b H / \nu = 5600$ where $U_b$ is the bulk velocity in the channel, $H$ the full channel  height and $\nu$ the kinematic viscosity of the fluid. An unstructured grid is used to resolve the trapezoidal riblets at hand. Figure \ref{Choi} shows the present grid in comparison to the one used by Choi et al. \cite{choi1993direct} more than 20 years ago. While those riblets have a tip angle of $\alpha=60^\circ$ the second investigated riblet geometry with a tip angle of $\alpha=53.5^\circ$ and larger spanwise spacing is shown in figure \ref{ribs}; as conventionally used in literature $h$ corresponds to the riblets height and $s$ to the riblet spacing. 


\begin{figure}[h!]
\centering
\includegraphics[width=0.45\textwidth]{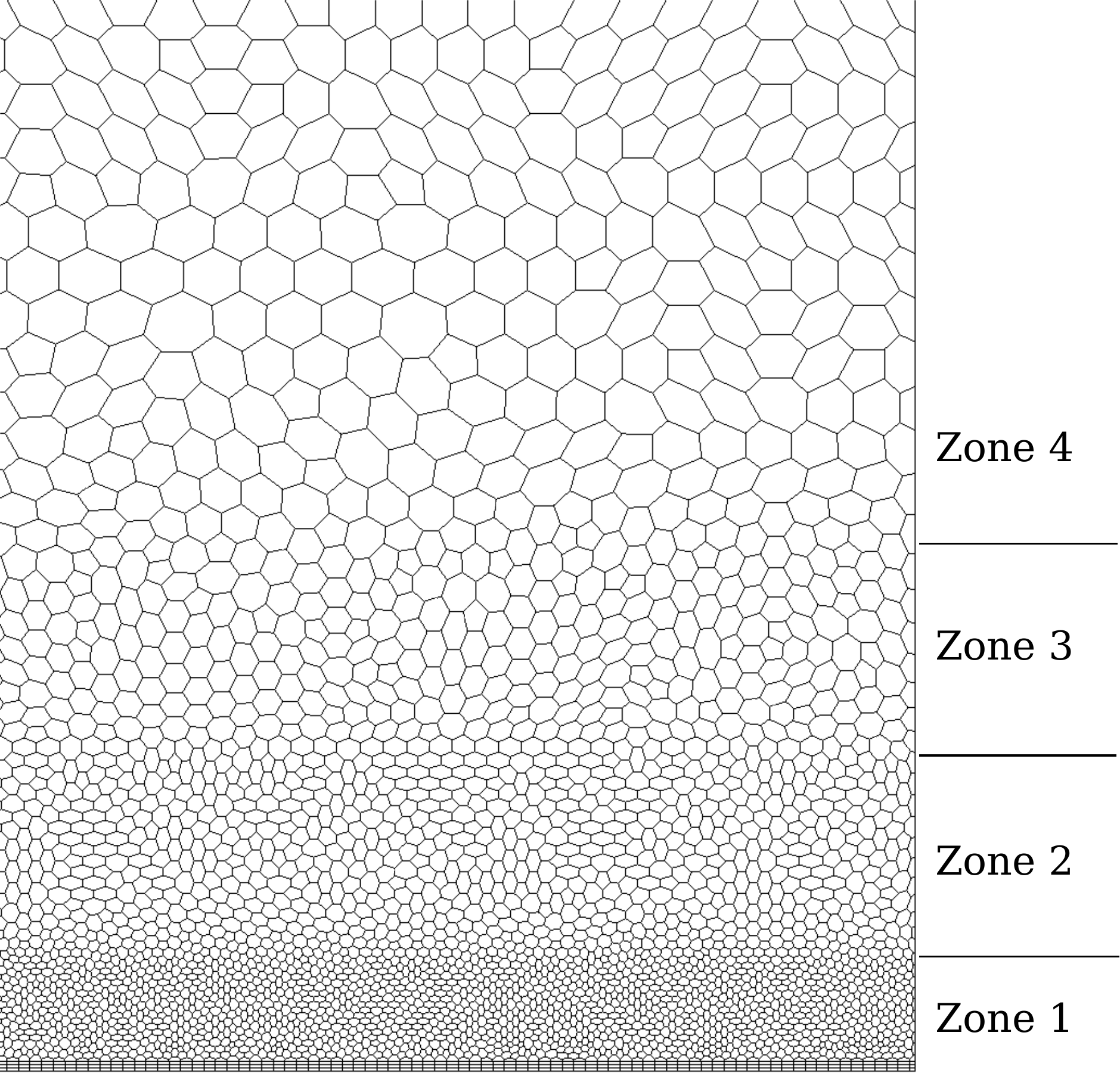}
\caption{Segment of the smooth channel resolved with unstructured polyhedral cells.}
\label{ZonenGlatt}
\end{figure}

\begin{figure}[h]
\centering
\includegraphics[width=1\textwidth]{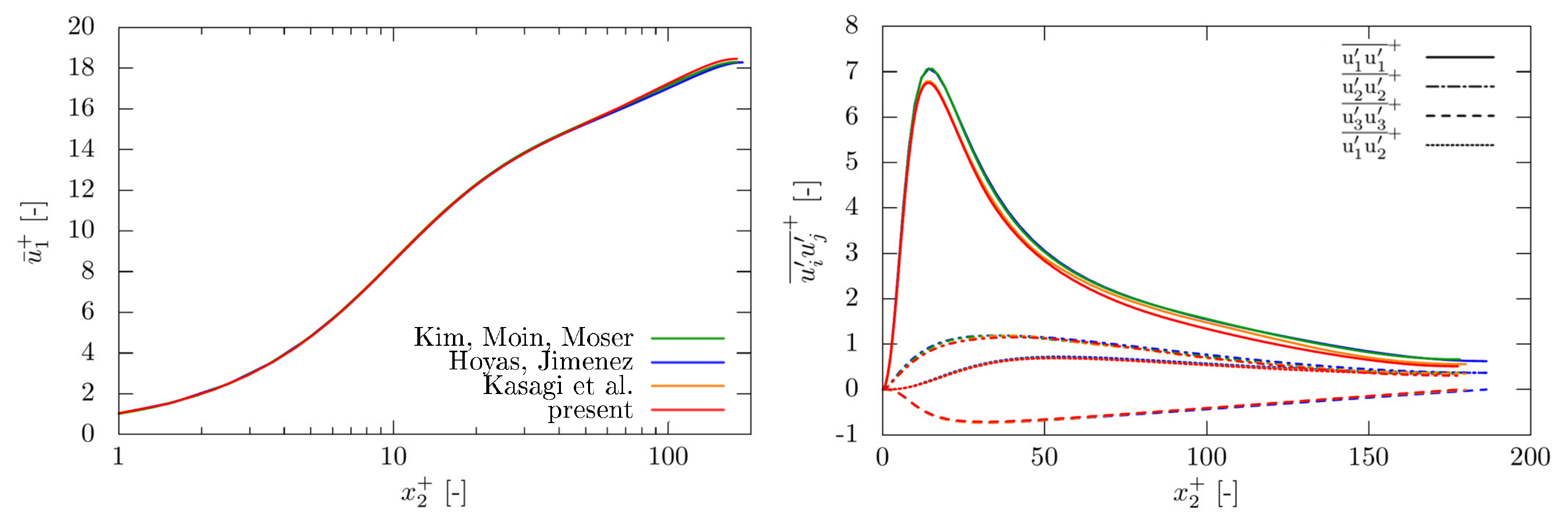}
\caption{Reference Case - smooth channel in comparison with literature data.}
\label{ValidationRefernce}
\end{figure}

In order to gain some understanding of the quality of the numerical procedure the reference case of a channel with smooth walls at  $Re_b= 5600$ ($Re_\tau = 0.5H u_\tau / \nu \approx 180$) is also computed on an unstructured grid (similar to the one used for the riblet simulations) as shown in figure \ref{ZonenGlatt}. Here, the spacial discretisation in streamwise direction ($x_1$) is constant while it varies in wall-normal ($x_2$) and spanwise ($x_3$) direction. Further details on the numerical set-up in  OpenFOAM$\tiny{^{\textregistered}}$ can be found in \cite{fink, daschiel2013numerical}. The obtained results are in good agreement with literature data (see table \ref{tab:RefSmooth}) as shown in figure \ref{ValidationRefernce}. Note that the peak value for  $\overline{u_1'u_1'}$ basically coincides with the finite difference based dataset \cite{kasa} while there are small differences in comparison to spectral discretization schemes \cite{moser, jim}.

\begin{table}[h]
\centering
\begin{tabular}{| c | c | c |}
\hline
\small{Case} & \small{$\mathrm{Re_{\tau}}$} & \small{Discretization}\\
\hline
\small{Kim, Moin, Moser \cite{moser1999dns, moser}} & \small{$\approx$ 180} & \small{spectral} \\
\hline
\small{Hoyas, Jimenez \cite{hoyas, jim}} & \small{$\approx$ 180} & \small{spectral} \\
\hline
\small{Kasagi et al. \cite{kasa}} & \small{$\approx$ 180} & \small{finite differences} \\
\hline
\end{tabular}
\caption{Literature references for the present smooth channel.}
\label{tab:RefSmooth}
\end{table}

%

The set-up of the present DNS are summerized in table \ref{tab:Sims}. Note that the case with  $\alpha=60^\circ$ is set up to resemble the study of Choi et al. \cite{choi1993direct}. Therefore, riblets are placed on one of the two channel walls only and drag reduction is evaluated through a comparison between this riblet wall with the smooth wall on the opposite side of the channel. The second riblet type is investigated in a similar manner as in the corresponding experimental study \cite{guttler2015high}: riblets are placed on the top and bottom of the channel and the resulting drag is compared to the reference channel with smooth walls at the same net channel height $H$. The same net channel height indicates that the internal volume of the channel with and without riblets is identical.

 In order to estimate the riblet size in viscous units for the prescribed bulk Reynolds number, we evaluate the friction velocity $u_\tau=\sqrt{\tau_w/ \rho}$ based on Dean's correlation \cite{dean1978reynolds}. The resulting friction based Reynolds number is $Re_\tau = \frac{u_\tau 0.5H}{\nu} \approx 180$. The viscous riblet dimensions $h^+$ and $s^+$ are chosen such that drag reduction is expected for case I whereas drag increase is expected for case II.


\begin{table}[h]
\centering
\begin{tabular}{| c | c | c | c | c | c | c |}
\hline
\small{Case} & \small{$\mathrm{Re_b}$}  & \small{$\mathrm{h^+}$} & \small{$\mathrm{s^+}$} & \small{Domainsize} & \small{$\mathrm{\Delta x_1^+}$}  & \small{$\mathrm{\Delta x_2^+/\Delta x_3^+}$}\\
\hline
\small{smooth} & \small{5600} & \small{-} & \small{-} & \small{$4 \pi \delta \times 2\delta \times 2/3 \pi \delta$} & \small{$\sim$ 9.3} & \small{$\sim$ 0.6...3.8} \\
\hline
\small{$\alpha=60^\circ$ \cite{choi1993direct}} & \small{5600} & \small{17.3} & \small{20} & \small{$\pi \delta \times 2\delta \times 0.289 \pi \delta$} & \small{$\sim$ 9.3} & \small{$\sim$ 0.6...3.8} \\
\hline
\small{$\alpha=53.5^\circ $ - case I} & \small{5600} & \small{8.4} & \small{17.3} & \small{$4 \pi \delta \times 2\delta \times 2/3 \pi \delta$} & \small{$\sim$ 9.3} & \small{$\sim$ 0.6...3.8} \\
\hline
\small{$\alpha=53.5^\circ $ - case II} & \small{5600} & \small{14.5} & \small{30} & \small{$4 \pi \delta \times 2\delta \times 2/3 \pi \delta$} & \small{$\sim$ 9.3} & \small{$\sim$ 0.6...3.8} \\
\hline
\end{tabular}
\caption{Numerical set-up of the investigated cases: All DNS are run at a constant bulk Reynolds number of $Re_b=5600$, the corresponding dimensionless sizes of the investigated riblets are given by $h^+$ and $s^+$. The size of the numerical domain is described in multiples of the channel half height $\delta = 0.5H$ and the grid resolution in viscous units is given for all three spatial directions by $\Delta x_i^+$.}
\label{tab:Sims}
\end{table}

\section{Results}
\label{sec3}

For DNS at constant flow rate drag reduction $R$ is defined as the relative difference between the average wall shear stress on the smooth wall ($\tau_{w,0}$) and on the riblet channel wall ($\tau_{w,rib}$) 
\begin{equation}
R=\frac{\tau_{w,0}-\tau_{w,rib}}{\tau_{w,0}}.
\end{equation} 
For riblets with  $\alpha=60^\circ$ the present simulations result in $R=4.4\%$ which is below the value of $R=6\%$ reported in \cite{choi1993direct}. Note that in both cases the evaluation of $R$ is based on the data obtained in an asymmetric channel flow set-up, in which one smooth wall and one riblet wall are present. Experimentally obtained values for similar riblet geometries are $R=4.5\%$ by Bechert et al. \cite{bechert1997experiments} or $R=4\%$ by Walsh \cite{walsh1982turbulent} for riblets with $h^+ = s^+$, i.e. $\alpha \sim 53^\circ$.

\begin{figure}[h]
\centering
\includegraphics[width=0.6\textwidth]{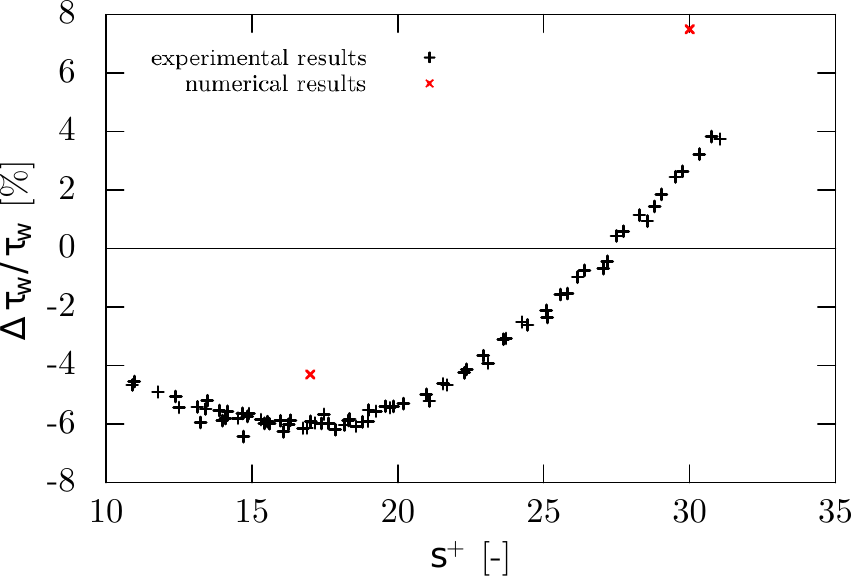}
\caption{Experimental (in black) and numerical (in red) results of the investigated trapezodial riblets.}
\label{RibletsExpNum}
\end{figure}

For the riblets depicted in figure \ref{ribs} ($\alpha=53.5^\circ $ with $s^+>h^+$) drag reduction is computed based on the comparision of $\tau_w$ in a symmetric riblet channel with a symmetric smooth wall channel at the same $H$ and $Re_b$. 
The resulting drag reduction is shown in figure \ref{RibletsExpNum} in comparison to the experimental results that were obtained with the same procedure. The results are plotted in terms of 
\begin{equation}
\Delta \tau_w/\tau_w=\frac{\tau_{w,rib}-\tau_{w,0}}{\tau_{w,0}}
\end{equation}
which corresponds to the notation typically employed in the presentation of the results from Bechert and co-workers. Here a negative $\Delta \tau_w$ indicates drag reduction. Figure \ref{RibletsExpNum} shows in black the results of the experimental investigation and in red the two DNS results. The DNS reveals the right trends: as expected $s^+=17.3$ results in drag reduction and $s^+=30$ results in drag increase. However, the actual values of the difference of wall shear stress differ between experiments and simulations. To get a better understanding for the accuracy of the actual drag reduction values, table \ref{tab:Rib} includes the related numerical and experimental error bars. Here, the numerical error bars were calculated according to \cite{oliver2014estimating}. Detailed information about the experimental error bars can be found in \cite{guttler2015high}. Note that the uncertainty in the numerical computation of $R$ is acutally higher than in the experiment. The results show that even with consideration of the error bars the obtained values for $R$ substantially differ between experiments and simulation.

\begin{table}[h]
\centering
\begin{tabular}{| c | c | c | c | c |}
\hline
\small{$\mathrm{s^+}$} & \small{R numerical} & \small{$Re_b$ numerical} & \small{R experimental} & \small{$Re_b$ experimental}\\
\hline
\small{17.3} & \small{4.3\% $+1.1\% \atop -1.1\%$} & \small{5600} & \small{6.1\% $+0.4\% \atop -0.4\%$} & \small{12500}\\
\hline
\small{30} & \small{-7.5\% $+1.4\% \atop -1.4\%$} & \small{5600} & \small{-2.8\% $+0.2\% \atop -0.2\%$} & \small{22000}\\
\hline
\end{tabular}
\caption{Details of numerical and experimental results for $s^+=17.3$ and $s^+=30$.}
\label{tab:Rib}
\end{table}

A potential reason for this discrepancy lies in the fact that the investigations are not carried out at the same bulk Reynolds number. 
Table \ref{tab:Rib} also shows the respective bulk Reynolds numbers of the numerical and experimental studies. The bulk Reynolds number of the experiment is a factor $2-4$ larger than in the DNS. While riblets drag reduction is classically believed to scale in visous units it should be noted that at the very low Reynolds number of the DNS, the riblets are relatively large compared to the channel height such that their influence in terms of a spanwise modulation of the mean flow field is felt up to a large distance from the wall. 

\begin{figure}[h]
\centering
\includegraphics[width=1\textwidth]{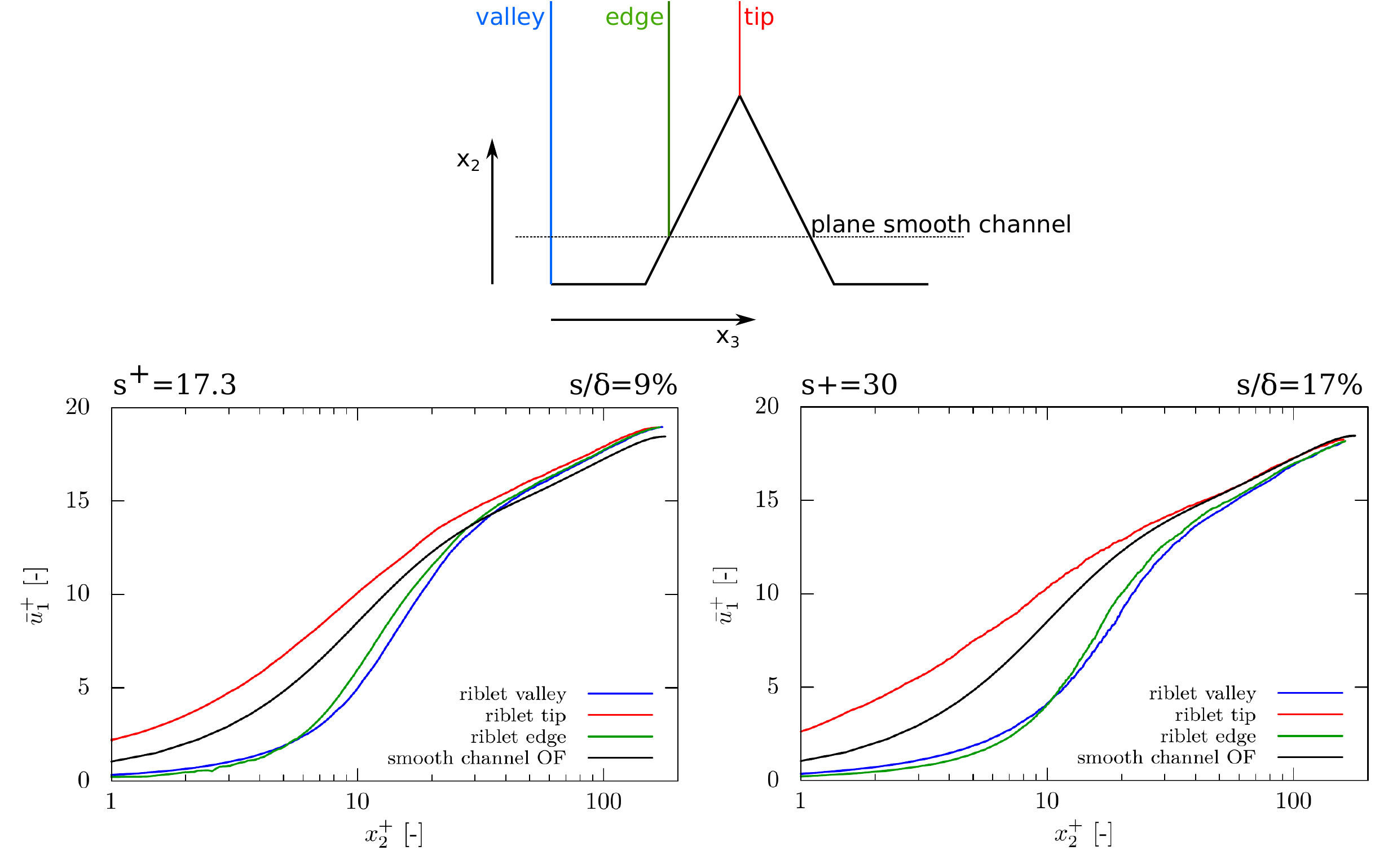}
\caption{Mean streamwise velocity component at different spanwise positions over the half channel height.}
\label{UmenaRibletsPosition}
\end{figure}

The mean velocity profile along selected locations above the riblets is shown in figure \ref{UmenaRibletsPosition}. Here, the distance from the wall corresponds to the distance from the reference position of the plane smooth channel as indicated in the top part of the figure. For the normalization in the smooth channel the friction velocity of the smooth channel is used. For the riblet cases the friction velocity from the corresponding riblet cases is used. The plots clearly show spanwise modulation of the velocity profiles. In the drag increasing case, spanwise differences are found all the way up to the center of the channel. 

\section*{Acknowledgement} 
Support through project FR2823/2 of the German Research Foundation (DFG) is greatly acknowledged.

\newpage

\bibliographystyle{elsarticle-num}

\bibliography{riblets}

\begin{thebibliography}{10}
\expandafter\ifx\csname url\endcsname\relax
  \def\url#1{\texttt{#1}}\fi
\expandafter\ifx\csname urlprefix\endcsname\relax\def\urlprefix{URL }\fi
\expandafter\ifx\csname href\endcsname\relax
  \def\href#1#2{#2} \def\path#1{#1}\fi

\bibitem{finkexperimental}
V.~Fink, A.~G{\"u}ttler, B.~Frohnapfel, Experimental and numerical
  investigation of riblets in a fully developed turbulent channel flow,
  European Drag Reduction and Flow Control Meeting EDRFCM 2015, Cambridge, UK.

\bibitem{choi1993direct}
H.~Choi, P.~Moin, J.~Kim, Direct numerical simulation of turbulent flow over
  riblets, Journal of Fluid Mechanics 255 (1993) 503--539.

\bibitem{guttler2015high}
A.~G{\"u}ttler, High accuracy determination of skin friction differences in an
  air channel flow based on pressure drop measurements, Ph.D. thesis,
  Karlsruhe, Karlsruher Institut f{\"u}r Technologie (KIT), Diss., 2015 (2015).

\bibitem{bechert1997experiments}
D.~Bechert, M.~Bruse, W.~Hage, J.~T. Van~der Hoeven, G.~Hoppe, Experiments on
  drag-reducing surfaces and their optimization with an adjustable geometry,
  Journal of Fluid Mechanics 338 (1997) 59--87.

\bibitem{fink}
V.~Fink, Numerische {U}ntersuchung der {B}eeinflussung der turbulenten
  {K}analstr{\"o}mung durch {R}ibelts, Master's thesis, KIT, Institut f{\"u}r
  Str{\"o}mungsmechanik (2014).

\bibitem{daschiel2013numerical}
G.~Daschiel, B.~Frohnapfel, J.~Jovanovi{\'c}, Numerical investigation of flow
  through a triangular duct: The coexistence of laminar and turbulent flow,
  International Journal of Heat and Fluid Flow 41 (2013) 27--33.

\bibitem{kasa}
http://thtlab.jp (2004).

\bibitem{moser}
http://turbulence.ices.utexas.edu/data/{MKM}/chan180 (2000).

\bibitem{jim}
http://torroja.dmt.upm.es/channels/data/fields/re180 (2013).

\bibitem{moser1999dns}
R.~Moser, J.~Kim, N.~Mansour, Dns of turbulent channel flow up to {R}e$_\tau$=
  590, Physics of Fluids 11 (1999) 943--945.

\bibitem{hoyas}
S.~Hoyas, J.~Jim{\'e}nez, Reynolds number effects on the reynolds-stress
  budgets in turbulent channels, Physics of Fluids 20 (2008) 101511.

\bibitem{dean1978reynolds}
R.~Dean, Reynolds number dependence of skin friction and other bulk flow
  variables in two-dimensional rectangular duct flow, J. Fluids Eng 100~(2)
  (1978) 215--223.

\bibitem{walsh1982turbulent}
M.~Walsh, Turbulent boundary layer drag reduction using riblets, in: 20th
  aerospace sciences meeting, 1982, p. 169.

\bibitem{oliver2014estimating}
T.~Oliver, N.~Malaya, R.~Ulerich, R.~Moser, Estimating uncertainties in
  statistics computed from direct numerical simulation, Physics of Fluids
  26~(3) (2014) 035101.

\end{thebibliography}

\end{document}